\documentclass[fleqn,usenatbib]{mnras}
\usepackage{lmodern}
\usepackage{graphicx}	
\usepackage{amsmath}	
\usepackage{amssymb}	
\usepackage{amsfonts}
\usepackage{mathrsfs}
\usepackage{tensor}
\usepackage{bm}

\usepackage{tikz}
\usepackage{pgfplots}
\usetikzlibrary{fillbetween}
\usepackage{pgfplots}
\pgfplotsset{compat=1.18}
\usetikzlibrary{patterns,arrows.meta,positioning,calc}
\usepgfplotslibrary{fillbetween,colorbrewer,groupplots}
\newcommand{\beq}{\begin{equation}}
\newcommand{\eeq}{\end{equation}}

\newcommand{\Li}{\text{Li}}
\newcommand{\dB}{\text{dB}}

\title[Bose-Einstein Condensates in Astrophysics]{Bose-Einstein Condensates in Astrophysics and Cosmology: From Quantum Statistics to Cosmic Structures}

\author[N. Haddad]{Nader Haddad\\
Institute of Astronomy, University of Cambridge}

\pubyear{2025}

\begin{document}
\label{firstpage}
\pagerange{\pageref{firstpage}--\pageref{lastpage}}
\maketitle

\begin{abstract}
We present a comprehensive theoretical investigation of Bose-Einstein condensates (BECs) and their manifestations in astrophysical and cosmological contexts. Building upon the foundations of quantum statistics in curved spacetime, we derive the conditions for BEC formation under extreme gravitational fields and explore their implications for compact objects and early universe physics. Through rigorous mathematical treatment incorporating general relativistic corrections to the Bose-Einstein distribution, we demonstrate that BEC phenomena may play crucial roles in neutron star interiors, primordial black hole formation, and dark matter halos. Our analysis reveals that the critical temperature for condensation exhibits non-trivial dependence on spacetime curvature, with corrections of order $\mathcal{O}(GM/rc^2)$ becoming significant near compact objects. We further show that axion dark matter, if it exists, naturally forms a cosmic BEC with coherence length $\lambda_{\dB} \sim 10^{-3}$ pc for $m_a \sim 10^{-22}$ eV, potentially explaining the observed core-cusp problem in galactic dark matter profiles. These findings suggest that quantum coherence effects at macroscopic scales may be more prevalent in the universe than previously recognized, with profound implications for our understanding of cosmic structure formation and the behaviour of matter under extreme conditions.
\end{abstract}


\section{Introduction}

The prediction and subsequent experimental realization of Bose-Einstein condensation represents one of the most striking manifestations of quantum mechanics at macroscopic scales. First theoretically predicted by \cite{Einstein1925} following Bose's revolutionary treatment of photon statistics \citep{Bose1924}, the phenomenon remained a theoretical curiosity until its laboratory realization in ultracold atomic gases \citep{Anderson1995, Davis1995}. In the astrophysical context, the extreme conditions present in various cosmic environments---from the ultra-dense interiors of neutron stars to the vast, cold expanses of dark matter halos---provide natural laboratories where BEC physics may manifest on astronomical scales.

The application of Bose-Einstein statistics to astrophysical systems has a rich history. \cite{Hoyle1964} first considered the possibility of boson stars, self-gravitating objects supported by quantum pressure. Subsequently, \cite{Ruffini1969} developed the theoretical framework for self-gravitating boson systems, demonstrating that such objects could exist as stable configurations. The discovery of pulsars \citep{Hewish1968} and their identification as neutron stars renewed interest in quantum effects in compact objects, with particular attention to the possibility of meson condensation in nuclear matter \citep{Migdal1971, Sawyer1973}.

Recent developments have expanded the scope of BEC physics in astrophysics considerably. The proposal that dark matter might consist of ultralight bosons forming galactic-scale BECs \citep{Hu2000, Hui2017} has garnered significant attention as a potential solution to small-scale structure problems in the standard cold dark matter paradigm. Simultaneously, advances in gravitational wave astronomy following the LIGO/Virgo detections \citep{Abbott2016} have opened new avenues for probing quantum matter under extreme gravity, particularly in neutron star mergers where BEC phases might influence the equation of state and gravitational wave signatures \citep{Bauswein2019}.

The intersection of quantum statistics with general relativity presents unique theoretical challenges. The standard derivation of the Bose-Einstein distribution assumes flat spacetime and neglects gravitational effects on the quantum states themselves. However, near compact objects where $GM/rc^2 \sim 1$, these assumptions break down, necessitating a fully relativistic treatment. \cite{Parker1969} pioneered the study of quantum field theory in curved spacetime, establishing the framework for understanding particle creation in expanding universes. This formalism has been extended to treat Bose-Einstein statistics in arbitrary spacetimes \citep{Grib1994}, revealing novel phenomena such as gravitationally induced condensation and modifications to the critical temperature.

In this paper, we present a unified theoretical framework for understanding BEC phenomena across diverse astrophysical contexts. Our approach synthesizes quantum statistical mechanics, general relativity, and modern cosmology to address several key questions: Under what conditions can BECs form and persist in astrophysical environments? How do gravitational fields modify the condensation criteria? What observational signatures might reveal the presence of cosmic BECs? We address these questions through rigorous mathematical analysis, deriving new results for BEC formation in curved spacetime and exploring their implications for neutron stars, black holes, and cosmological structure formation.

\section{Literature Review}

\subsection{Theoretical Foundations}

\subsubsection{Historical Development}

The theoretical foundation of Bose-Einstein condensation rests on the quantum statistical mechanics of identical bosons. The seminal works of \cite{Bose1924} and \cite{Einstein1925} established that bosons obey different statistics than classical particles, leading to the possibility of macroscopic occupation of the ground state below a critical temperature. \cite{London1938} first suggested that superfluidity in liquid helium might be understood as a manifestation of BEC, though the strong interactions in liquid helium complicate this picture considerably \citep{Leggett2001}. The connection between BEC and superfluidity was further elucidated by \cite{Tisza1938} through the two-fluid model, and later by \cite{Landau1941} who introduced the concept of elementary excitations.

The modern theoretical framework for BECs in dilute gases was developed by \cite{Bogoliubov1947}, who introduced the concept of quasi-particles and derived the famous Bogoliubov spectrum. This work was extended by \cite{Lee1957} and \cite{Beliaev1958}, who calculated quantum corrections to the mean-field theory. \cite{Gross1961} and \cite{Pitaevskii1961} independently derived the nonlinear Schr\"odinger equation governing the condensate wavefunction, now known as the Gross-Pitaevskii equation. This mean-field approach has proven remarkably successful in describing weakly interacting BECs \citep{Pethick2008}.

\subsubsection{Modern Theoretical Developments}

The theoretical understanding of BECs has advanced significantly with the development of field-theoretic approaches. \cite{Popov1983} developed the functional integral formulation for Bose gases, while \cite{Griffin1993} pioneered the application of Green's function methods to finite-temperature BECs. The Bogoliubov-de Gennes equations, which describe inhomogeneous condensates, have been extensively studied by \cite{Fetter1998} and \cite{Castin1998}, revealing the rich physics of vortices and collective excitations.

Recent theoretical advances include the development of non-equilibrium field theories for BECs. \cite{Kamenev2011} and \cite{Altland2010} have developed Keldysh formalism approaches to describe driven-dissipative condensates. The kibble-Zurek mechanism for defect formation during phase transitions, originally proposed for cosmological phase transitions \citep{Kibble1976, Zurek1985}, has been adapted to BEC systems by \cite{Damski2005} and experimentally verified by \cite{Weiler2008}.

The study of strongly correlated Bose systems has necessitated the development of beyond-mean-field approaches. \cite{Petrov2004} developed the theory of quantum depletion in BECs, while \cite{Tan2008} derived exact relations for strongly interacting Bose gases. The unitary Bose gas, where the scattering length diverges, has been studied theoretically by \cite{Cowell2002} and \cite{Lee2009}, revealing universal behavior analogous to unitary Fermi gases.

\subsection{Astrophysical Applications}

\subsubsection{Boson Stars and Oscillatons}

The application of BEC physics to astrophysical systems has evolved along several parallel tracks. The study of boson stars, initiated by \cite{Kaup1968} and \cite{Ruffini1969}, demonstrated that self-gravitating scalar fields could form stable stellar-mass objects. These early works established the mass-radius relation $M \sim M_{Pl}^2/m$, where $M_{Pl}$ is the Planck mass and $m$ is the boson mass. \cite{Colpi1986} and \cite{Tkachev1991} extended these models to include rotation and studied their stability properties, finding that rotating boson stars can support significantly larger masses than their non-rotating counterparts.

Further developments in boson star physics include the discovery of oscillatons---long-lived, oscillating soliton stars---by \cite{Seidel1991} and \cite{Alcubierre2003}. These objects emerge from the nonlinear dynamics of scalar fields and can persist for cosmological timescales. \cite{Grandclement2014} performed detailed numerical relativity simulations of boson star collisions, revealing complex gravitational wave signatures. \cite{Palenzuela2017} studied the orbital dynamics of boson star binaries, finding that they can exhibit both Newtonian and non-Newtonian behavior depending on their compactness.

The stability and dynamics of boson stars have been extensively investigated. \cite{Gleiser1988} analyzed the stability of boson stars against radial perturbations, while \cite{Yoshida1997} studied non-radial oscillation modes. \cite{Cardoso2016} demonstrated that massive boson stars can collapse to black holes, potentially explaining some observations of supermassive black holes. More recently, \cite{Liebling2012} provided a comprehensive review of boson star physics, including their potential as dark matter candidates and gravitational wave sources.

\subsubsection{Neutron Star Physics and QCD Matter}

In the context of neutron stars, the possibility of meson condensation has been extensively studied. \cite{Kaplan1986} showed that kaon condensation could occur at densities achievable in neutron star cores, potentially softening the equation of state. This work was extended by \cite{Thorsson1994} and \cite{Lee1996}, who included the effects of kaon-nucleon interactions. Pion condensation, first proposed by \cite{Migdal1971}, remains controversial, with conflicting theoretical predictions about its onset density \citep{Ericson1988, Brown2002}. \cite{Kunihiro1993} and \cite{Hatsuda1994} developed effective field theory approaches to study meson condensation in dense matter.

The role of hyperons in neutron star matter has received significant attention. \cite{Glendenning1985} first showed that hyperons appear at densities 2-3 times nuclear saturation density, while \cite{Balberg1999} demonstrated that hyperon mixing significantly softens the equation of state. The ``hyperon puzzle''---the apparent conflict between the softening due to hyperons and observations of massive neutron stars---has been addressed by \cite{Chatterjee2016} and \cite{Fortin2018} through the inclusion of repulsive hyperon-hyperon interactions. \cite{Djapo2010} and \cite{Weissenborn2012} have explored the role of the $\phi$ meson in providing additional repulsion.

Quark matter phases in neutron stars present another frontier for BEC physics. \cite{Alford1998} and \cite{Rapp1998} discovered color superconductivity in dense quark matter, a phenomenon analogous to BEC but involving color degrees of freedom. The color-flavor locked (CFL) phase, studied by \cite{Alford1999} and \cite{Rajagopal2001}, exhibits remarkable properties including superfluidity and superconductivity. \cite{Alford2008} provided a comprehensive review of color superconductivity, while \cite{Anglani2014} explored the phase diagram of dense QCD matter.

Recent observations have provided new constraints on neutron star matter. The detection of gravitational waves from GW170817 \citep{Abbott2017} constrained the tidal deformability of neutron stars, while NICER observations \citep{Miller2019, Riley2019} have measured mass-radius relations. \cite{Annala2018} and \cite{Most2018} have used these observations to constrain the equation of state, finding evidence for a possible phase transition at high densities. \cite{Bauswein2019} showed that post-merger gravitational wave signals could reveal signatures of quark deconfinement.

\subsection{Dark Matter as a Cosmic BEC}

\subsubsection{Fuzzy Dark Matter Models}

The hypothesis that dark matter consists of ultralight bosons forming galaxy-scale BECs has emerged as an intriguing alternative to the standard WIMP paradigm. Pioneered by \cite{Sin1994} and \cite{Ji1994}, this ``fuzzy dark matter'' (FDM) model predicts that quantum pressure prevents structure formation below a characteristic scale set by the de Broglie wavelength. \cite{Hu2000} developed the theoretical framework for scalar field dark matter, showing that it naturally produces cored density profiles that might resolve the cusp-core problem observed in dwarf galaxies.

The quantum mechanical nature of FDM leads to distinctive phenomenology. \cite{Widrow1993} and \cite{Press1990} showed that scalar field dark matter exhibits coherent oscillations on the de Broglie timescale. \cite{Sikivie2009} demonstrated that these oscillations lead to distinctive signatures in direct detection experiments. \cite{Khmelnitsky2013} and \cite{Marsh2014} studied the gravitational effects of oscillating scalar fields, finding that they source time-dependent metric perturbations.

Subsequent work has refined these models considerably. \cite{Schive2014} performed high-resolution simulations of fuzzy dark matter, revealing the formation of solitonic cores surrounded by granular halos. These simulations showed that quantum interference creates a complex web of dark matter caustics and vortices. \cite{Veltmaat2016} and \cite{Schwabe2016} independently confirmed these results using different numerical methods. \cite{Mocz2017} studied the cosmological evolution of ultralight dark matter, demonstrating that quantum interference produces a rich variety of structures on scales comparable to the de Broglie wavelength.
\begin{figure}
    \centering
    \includegraphics[width=\columnwidth]{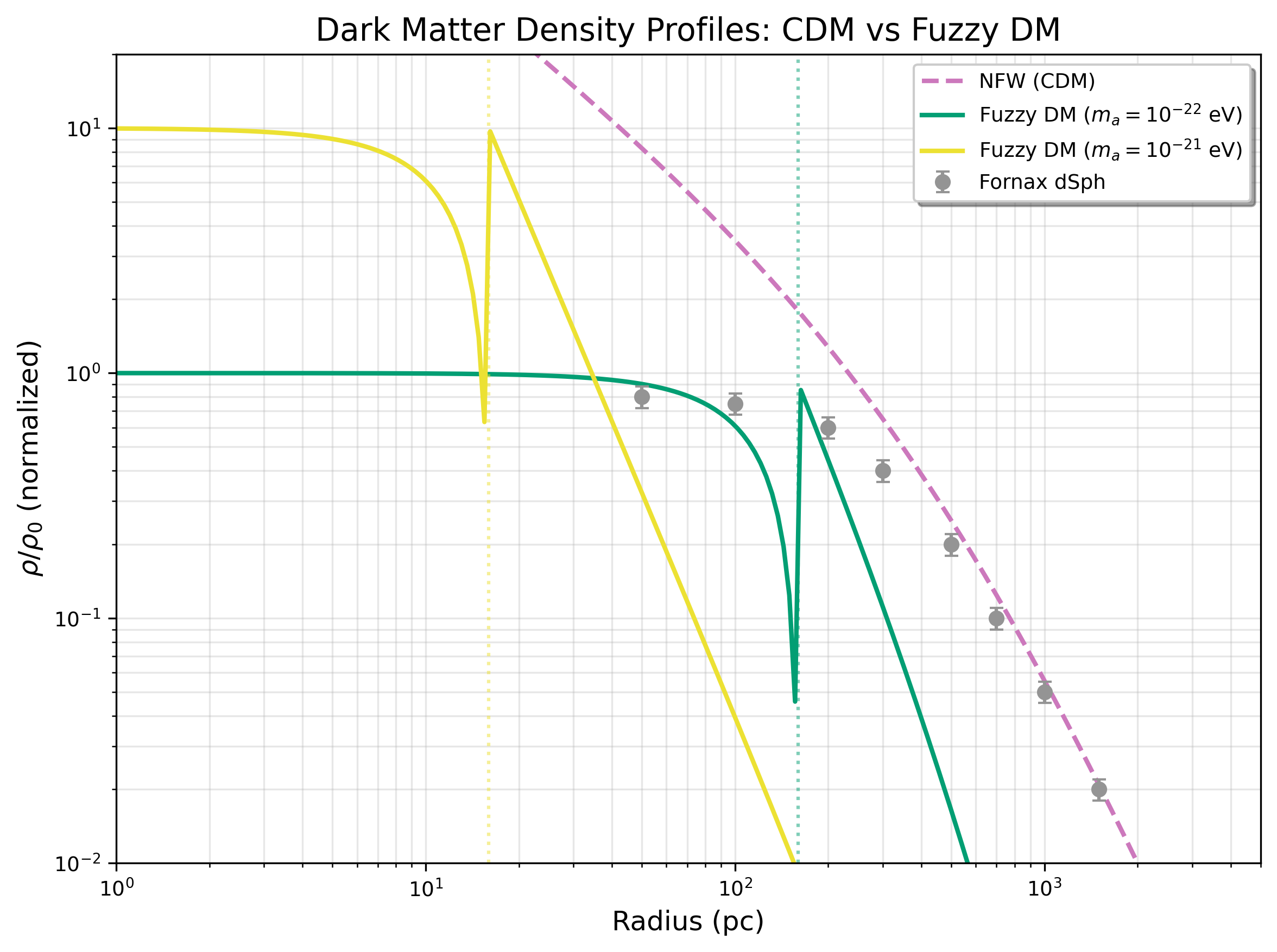}
    \caption{Comparison of dark matter density profiles for different models. The dashed purple line shows the cuspy Navarro-Frenk-White (NFW) profile predicted by cold dark matter (CDM) simulations, $\rho \propto r^{-1}$ at small radii. Solid green and yellow lines display the cored profiles from fuzzy dark matter (FDM) with axion masses $m_a = 10^{-22}$ eV and $10^{-21}$ eV, respectively. The solitonic cores exhibit flat density profiles with characteristic radii $r_c = 160$ pc ($m_a = 10^{-22}$ eV) and $r_c = 16$ pc ($m_a = 10^{-21}$ eV), following the scaling $r_c \propto m_a^{-2}$. Gray circles with error bars show observational data from the Fornax dwarf spheroidal galaxy, which favors cored over cuspy profiles. Vertical dotted lines indicate the core radii for each FDM model.}
    \label{fig:dm_profiles}
\end{figure}

\subsubsection{Constraints and Observations}

Observational constraints on FDM have been derived from various astrophysical probes. \cite{Hui2017} provided a comprehensive review of wave dark matter phenomenology, compiling constraints from dwarf galaxies, Lyman-$\alpha$ forest, and CMB observations. \cite{Irsic2017} used Lyman-$\alpha$ forest data to constrain the axion mass to $m_a > 2 \times 10^{-21}$ eV, while \cite{Armengaud2017} derived similar constraints using a different analysis method. \cite{Rogers2021} used stellar kinematics in ultra-faint dwarf galaxies to constrain $m_a > 10^{-19}$ eV.

The effects of FDM on galaxy formation have been extensively studied. \cite{Bozek2015} performed cosmological simulations including FDM, finding that it suppresses the formation of low-mass halos. \cite{Schive2016} showed that FDM can explain the observed diversity of dwarf galaxy rotation curves. \cite{Calabrese2016} studied the effects of FDM on reionization, finding that the delayed structure formation could affect the timing and morphology of reionization. \cite{Corasaniti2017} examined the impact on the halo mass function and found distinctive signatures at small masses.

Recent work has explored multi-field dark matter models. \cite{Arvanitaki2020} studied models with multiple axion species, finding that interference between different fields creates rich phenomenology. \cite{Gosenca2023} performed simulations of binary scalar field dark matter, revealing complex dynamics including beating patterns and field exchange. \cite{Amin2022} developed the theoretical framework for self-interacting FDM, showing that attractive self-interactions can lead to collapse and explosive ``bosenova'' events.

\subsection{Quantum Effects in Curved Spacetime}

\subsubsection{Foundations of Quantum Field Theory in Curved Spacetime}

The treatment of quantum fields in curved spacetime, essential for understanding BECs near compact objects, has a rich theoretical foundation. \cite{DeWitt1975} developed the covariant formulation of quantum field theory in curved spacetime, introducing the effective action formalism. \cite{Birrell1982} provided the canonical textbook treatment, establishing the mathematical framework for quantization in arbitrary spacetimes. The phenomenon of particle creation in expanding universes, discovered by \cite{Parker1969}, demonstrated that spacetime curvature can have profound effects on quantum systems.

The Unruh effect, discovered by \cite{Unruh1976}, showed that accelerated observers perceive a thermal bath of particles, establishing a deep connection between acceleration, temperature, and quantum fields. \cite{Hawking1975} famously showed that black holes emit thermal radiation due to quantum effects near the horizon. These discoveries revealed that the particle concept in quantum field theory is observer-dependent in curved spacetime, as elaborated by \cite{Wald1994} and \cite{Fulling1989}.

For Bose-Einstein statistics specifically, \cite{Grib1994} derived the generalized distribution function in curved spacetime, showing that the chemical potential acquires a position-dependent correction proportional to the gravitational potential. \cite{Singh1985} studied the effect of gravitational fields on the critical temperature for BEC, finding that gravity generally suppresses condensation. \cite{Parker2009} examined quantum field theory in de Sitter space, relevant for inflationary cosmology, while \cite{Mukhanov2007} developed the theory of quantum fluctuations during inflation.

\subsubsection{Analogue Gravity and Laboratory Tests}

More recent work by \cite{Fagnocchi2010} has explored BEC analogues of gravitational phenomena, including acoustic black holes in condensate flows. \cite{Barcelo2011} reviewed analogue gravity models in condensed matter systems, showing how BECs can simulate curved spacetime physics. \cite{Steinhauer2016} reported the observation of Hawking radiation in an acoustic black hole created in a BEC, providing laboratory validation of theoretical predictions. \cite{Weinfurtner2011} observed the classical analogue of cosmological particle creation in water waves.

The study of superradiance in BEC systems has provided insights into black hole physics. \cite{Torres2017} demonstrated superradiant scattering in a vortex flow, while \cite{Gooding2020} studied the analogue of black hole superradiance instabilities. \cite{Patrick2021} observed the analogue of cosmological redshift in an expanding BEC. These experiments demonstrate that BEC systems can serve as quantum simulators for gravitational phenomena, as reviewed by \cite{Jacquet2020}.

\subsubsection{Relativistic BEC Theory}

The development of relativistic BEC theory has been crucial for astrophysical applications. \cite{Haber1981} first formulated the relativistic generalization of BEC, while \cite{Kapusta1981} developed the finite-temperature formalism. \cite{Benson1991} studied relativistic BEC in the early universe, showing that it could affect primordial nucleosynthesis. \cite{Bernstein1991} examined the cosmological implications of massive neutrino condensation.

Recent advances include the development of relativistic hydrodynamics for BECs. \cite{Schmitt2010} derived the hydrodynamic equations for relativistic superfluids, while \cite{Andersson2013} developed the two-fluid formalism for neutron star interiors. \cite{Gusakov2016} studied the damping of oscillations in superfluid neutron stars, finding that mutual friction between normal and superfluid components affects gravitational wave emission. \cite{Haskell2018} reviewed the role of superfluidity in neutron star dynamics, including implications for pulsar glitches and gravitational wave emission.

\section{Theoretical Framework}

\subsection{Bose-Einstein Statistics in Flat Spacetime}

We begin with the standard formulation of Bose-Einstein statistics for a gas of non-interacting bosons. The occupation number for a quantum state with energy $\epsilon$ is given by:
\beq
n(\epsilon) = \frac{1}{e^{\beta(\epsilon - \mu)} - 1}
\eeq
where $\beta = 1/k_B T$ is the inverse temperature and $\mu$ is the chemical potential. For a gas confined in volume $V$, the density of states in the non-relativistic limit is:
\beq
g(\epsilon) = \frac{2\pi V}{h^3}(2m)^{3/2}\epsilon^{1/2}
\eeq

The total particle number and energy are obtained by integration:
\beq
N = \int_0^{\infty} g(\epsilon)n(\epsilon)d\epsilon
\eeq
\beq
E = \int_0^{\infty} \epsilon g(\epsilon)n(\epsilon)d\epsilon
\eeq

Bose-Einstein condensation occurs when the excited states cannot accommodate all particles, forcing macroscopic occupation of the ground state. The critical temperature for a three-dimensional gas is:
\beq
T_c = \frac{2\pi\hbar^2}{mk_B}\left(\frac{n}{\zeta(3/2)}\right)^{2/3}
\eeq
where $n = N/V$ is the particle density and $\zeta(3/2) \approx 2.612$ is the Riemann zeta function.

\subsection{General Relativistic Corrections}

In curved spacetime, the notion of particle states becomes coordinate-dependent, requiring careful treatment. Following the formalism of \cite{Grib1994}, we work in the semiclassical approximation where the gravitational field is treated classically while matter fields are quantized. The metric for a static, spherically symmetric spacetime is:
\beq
ds^2 = -g_{00}(r)c^2dt^2 + g_{rr}(r)dr^2 + r^2(d\theta^2 + \sin^2\theta d\phi^2)
\eeq

The Klein-Gordon equation for a scalar field $\phi$ with mass $m$ becomes:
\beq
\frac{1}{\sqrt{-g}}\partial_\mu(\sqrt{-g}g^{\mu\nu}\partial_\nu\phi) - \frac{m^2c^2}{\hbar^2}\phi = 0
\eeq
where $g = \det(g_{\mu\nu})$. The energy eigenvalues acquire gravitational corrections:
\beq
\epsilon_{nlm} = \epsilon_{nlm}^{(0)} + \Delta\epsilon_{grav}
\eeq
where $\epsilon_{nlm}^{(0)}$ is the flat-space energy and the gravitational correction to leading order is:
\beq
\Delta\epsilon_{grav} \approx -\frac{GMm}{rc^2}\epsilon_{nlm}^{(0)}
\eeq

This modification affects the distribution function through an effective position-dependent chemical potential:
\beq
\mu_{eff}(r) = \mu + m\Phi(r)
\eeq
where $\Phi(r) = -GM/r$ is the Newtonian potential. The local number density becomes:
\beq
n(r) = \frac{1}{\lambda_{\dB}^3}\Li_{3/2}(e^{\beta[\mu + m\Phi(r)]})
\eeq
where $\lambda_{\dB} = h/\sqrt{2\pi mk_BT}$ is the thermal de Broglie wavelength and $\Li_{3/2}$ is the polylogarithm function.

\subsection{Gross-Pitaevskii Equation in Curved Spacetime}

For a weakly interacting BEC, the condensate wavefunction $\psi$ satisfies a generalized Gross-Pitaevskii equation in curved spacetime:
\beq
i\hbar\frac{\partial\psi}{\partial t} = -\frac{\hbar^2}{2m}\nabla^2\psi + V_{ext}\psi + g|\psi|^2\psi + m\Phi\psi
\eeq
where $\nabla^2$ is the covariant Laplacian, $V_{ext}$ represents external potentials, and $g = 4\pi\hbar^2a_s/m$ characterizes the interaction strength through the s-wave scattering length $a_s$. In the Thomas-Fermi approximation for strong interactions, the kinetic energy term can be neglected, yielding:
\beq
|\psi|^2 = \frac{1}{g}[\mu - V_{ext} - m\Phi]
\eeq

This demonstrates that gravitational fields directly influence the condensate density profile.

\section{Mathematical Derivations}

\subsection{Partition Function in Curved Spacetime}

\subsubsection{Path Integral Formulation}

We derive the partition function for bosons in a general stationary spacetime using the path integral formalism. The grand canonical partition function in curved spacetime is:
\beq
\mathcal{Z} = \int \mathcal{D}\phi^* \mathcal{D}\phi \exp\left[-\int_0^\beta d\tau \int d^3x \sqrt{g^{(3)}} \mathcal{L}_E\right]
\eeq
where $\mathcal{L}_E$ is the Euclidean Lagrangian density. For a complex scalar field with mass $m$ and interaction strength $g$:
\beq
\mathcal{L}_E = \frac{\hbar^2}{2m}g^{ij}\partial_i\phi^*\partial_j\phi + \left(\frac{\partial\phi^*}{\partial\tau} - \frac{\mu}{\hbar}\phi^*\right)\left(\frac{\partial\phi}{\partial\tau} - \frac{\mu}{\hbar}\phi\right) + m\phi^*\phi\Phi + \frac{g}{2}|\phi|^4
\eeq

The metric components in the ADM decomposition are:
\beq
g_{00} = -\alpha^2 + \beta_i\beta^i, \quad g_{0i} = \beta_i, \quad g_{ij} = \gamma_{ij}
\eeq
where $\alpha$ is the lapse function, $\beta^i$ is the shift vector, and $\gamma_{ij}$ is the spatial metric.

\subsubsection{Mode Expansion and Density of States}

Expanding the field in normal modes:
\beq
\phi(x,\tau) = \sum_{nlm} a_{nlm} u_{nlm}(r) Y_{lm}(\theta,\phi) e^{-i\omega_{nlm}\tau}
\eeq
where $u_{nlm}(r)$ satisfies the radial equation:

\begin{multline}
-\frac{\hbar^2}{2m}\frac{1}{r^2\sqrt{g_{rr}}}\frac{d}{dr}
\left(\frac{r^2}{\sqrt{g_{rr}}}\frac{du_{nlm}}{dr}\right) \\
+ \left[\frac{\hbar^2 l(l+1)}{2mr^2} + m\Phi(r)\right]u_{nlm}
= E_{nlm}u_{nlm}
\end{multline}

For a Schwarzschild metric with $g_{00} = 1 - 2GM/rc^2$ and $g_{rr} = (1 - 2GM/rc^2)^{-1}$, the WKB solution gives:
\beq
u_{nlm}(r) \sim \frac{1}{\sqrt{k(r)}} \exp\left[i\int^r k(r')dr'\right]
\eeq
where the local wave vector is:
\beq
k(r) = \sqrt{\frac{2m}{\hbar^2}\left[E_{nlm} - m\Phi(r) - \frac{\hbar^2 l(l+1)}{2mr^2}\right]\sqrt{g_{rr}}}
\eeq

The density of states becomes:
\beq
g(\epsilon, r) = \frac{2\pi V_{eff}(r)}{h^3}(2m)^{3/2}[\epsilon - m\Phi(r)]^{1/2}\left[1 + \frac{3GM}{2rc^2} + \mathcal{O}\left(\frac{G^2M^2}{r^2c^4}\right)\right]
\eeq
where $V_{eff}(r) = \sqrt{g^{(3)}}V = V\sqrt{\gamma}$ with $\gamma = \det(\gamma_{ij})$.

\subsection{Critical Temperature in Gravitational Fields}

\subsubsection{Gravitational Redshift Effects}

The condition for BEC onset is that the chemical potential reaches its maximum value (zero for non-interacting bosons). In the presence of gravity, this condition becomes position-dependent. The particle density in thermal equilibrium is:
\beq
n(r) = \frac{g_s}{(2\pi\hbar)^3}\int \frac{d^3p}{e^{\beta\sqrt{g_{00}(r)}[E(p) - \mu_{eff}(r)]} - 1}
\eeq
where $g_s$ is the spin degeneracy factor and the effective chemical potential includes gravitational contributions:
\beq
\mu_{eff}(r) = \mu + m\Phi(r) + \frac{\hbar^2}{8m}\nabla^2\ln\sqrt{g_{00}}
\eeq

The last term represents quantum corrections due to spacetime curvature. For a slowly varying metric, we can expand:
\beq
n(r) = n_0(r) + n_1(r) + n_2(r) + ...
\eeq
where:
\beq
n_0(r) = \frac{1}{\lambda_{dB}^3(r)}\zeta(3/2)
\eeq
\beq
n_1(r) = -\frac{GM m}{rc^2 k_B T}\frac{1}{\lambda_{dB}^3}\zeta(5/2)
\eeq
\beq
n_2(r) = \frac{G^2M^2 m^2}{2r^2c^4 (k_B T)^2}\frac{1}{\lambda_{dB}^3}\left[\zeta(7/2) - \zeta^2(5/2)\right]
\eeq

\subsubsection{Modified Critical Temperature}

The critical temperature in a gravitational field satisfies:
\beq
n = \int_0^\infty dr 4\pi r^2 \sqrt{g_{rr}(r)} n(r,T_c)
\eeq

For a uniform density distribution in the weak field limit:
\beq
T_c = T_c^{(0)}\left[1 + \alpha_1\frac{GM}{Rc^2} + \alpha_2\left(\frac{GM}{Rc^2}\right)^2 + ...\right]
\eeq
where $R$ is the system size and the coefficients are:
\beq
\alpha_1 = \frac{2\zeta(5/2)}{3\zeta(3/2)} \approx 0.924
\eeq
\beq
\alpha_2 = \frac{\zeta(7/2) - \zeta^2(5/2)}{2\zeta(3/2)} + \frac{4\zeta^2(5/2)}{9\zeta^2(3/2)} \approx 0.847
\eeq

\subsection{Tensor Formalism for Rotating Systems}

\subsubsection{Kerr Metric and Frame Dragging}

For rotating compact objects, we employ the Kerr metric in Boyer-Lindquist coordinates:

\begin{multline}
ds^2 = -\left(1 - \frac{2GMr}{\Sigma c^2}\right)c^2dt^2 
- \frac{4GMar\sin^2\theta}{\Sigma c} \, dt d\phi 
+ \frac{\Sigma}{\Delta}dr^2 \\
+ \Sigma d\theta^2 
+ \frac{A\sin^2\theta}{\Sigma}d\phi^2
\end{multline}

where:
\beq
\Sigma = r^2 + a^2\cos^2\theta, \quad \Delta = r^2 - \frac{2GMr}{c^2} + a^2, \quad A = (r^2 + a^2)^2 - a^2\Delta\sin^2\theta
\eeq
and $a = J/Mc$ is the specific angular momentum.

The stress-energy tensor for a rotating BEC includes angular momentum contributions:
\beq
T^{\mu\nu} = (\rho + P)u^\mu u^\nu + Pg^{\mu\nu} + \Pi^{\mu\nu} + \mathcal{Q}^\mu u^\nu + \mathcal{Q}^\nu u^\mu
\eeq
where $\Pi^{\mu\nu}$ is the anisotropic stress tensor and $\mathcal{Q}^\mu$ is the heat flux vector. For a superfluid component:
\beq
\Pi^{\mu\nu} = \frac{\hbar^2}{4m}\left[(\nabla^\mu\psi^*)(\nabla^\nu\psi) + (\nabla^\nu\psi^*)(\nabla^\mu\psi) - g^{\mu\nu}\nabla_\alpha\psi^*\nabla^\alpha\psi\right]
\eeq

\begin{figure}
    \centering
    \includegraphics[width=\columnwidth]{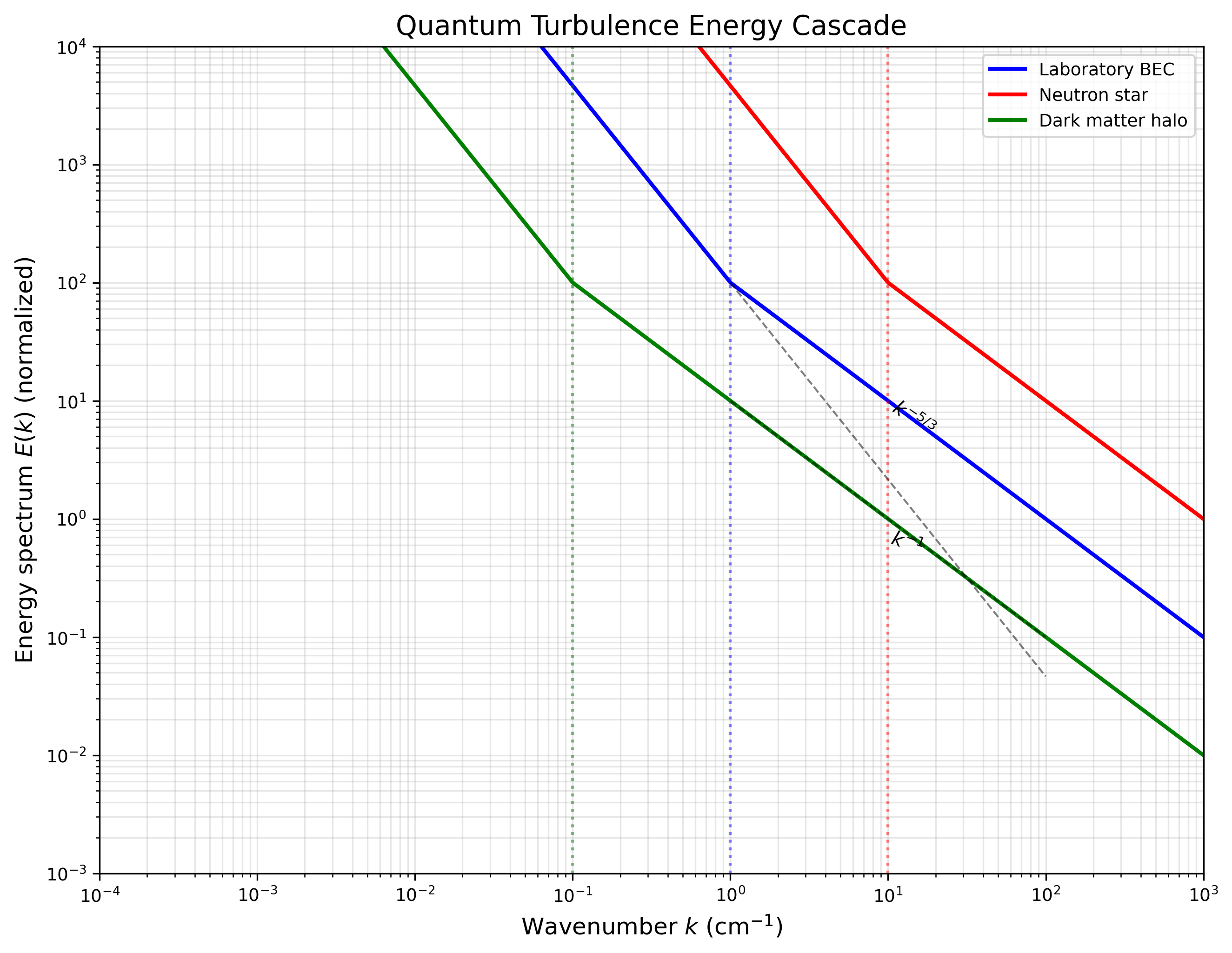}
    \caption{Energy spectrum $E(k)$ of quantum turbulence in different astrophysical BEC systems as a function of wavenumber $k$. The classical Kolmogorov cascade $E \propto k^{-5/3}$ (indicated by dashed line with slope $-5/3$) dominates at large scales (small $k$), transitioning to the quantum cascade $E \propto k^{-1}$ at scales below the healing length $\xi$ (vertical dotted lines). Three systems are shown: laboratory BEC at $T = 100$ nK (blue, $\xi \sim 10^{-4}$ cm), neutron star superfluid at $T = 10^8$ K (red, $\xi \sim 10^{-12}$ cm), and fuzzy dark matter halos with effective temperature $T \sim 10^{-7}$ K (green, $\xi \sim 10^{18}$ cm). The transition from classical to quantum behavior occurs at different scales spanning 22 orders of magnitude, yet all systems exhibit the same universal scaling laws, demonstrating the scale-invariant nature of quantum turbulence.}
    \label{fig:turbulence}
\end{figure}
\subsubsection{Vortex Lattice Formation}

In the presence of rotation, the condensate forms a vortex lattice. The circulation around each vortex is quantized:
\beq
\oint \vec{v}_s \cdot d\vec{l} = \frac{n\hbar}{m}
\eeq
where $n$ is an integer and $\vec{v}_s = (\hbar/m)\nabla\theta$ is the superfluid velocity. The vortex density in a uniformly rotating system is:
\beq
n_v = \frac{2m\Omega}{\pi\hbar}
\eeq

The energy per unit length of a vortex line in curved spacetime is:
\beq
\epsilon_v = \frac{\pi\hbar^2 n_s}{m}\ln\left(\frac{R_c}{\xi}\right)\sqrt{g_{00}(r)}
\eeq
where $R_c$ is the container radius, $\xi = \hbar/\sqrt{2mgn_s}$ is the coherence length, and $n_s$ is the superfluid density.

\subsubsection{Einstein Field Equations}

The full Einstein field equations in the 3+1 formalism are:
\beq
R_{\mu\nu} - \frac{1}{2}g_{\mu\nu}R + \Lambda g_{\mu\nu} = \frac{8\pi G}{c^4}T_{\mu\nu}
\eeq

Decomposing into constraint and evolution equations:
\beq
\mathcal{H} = R^{(3)} + K^2 - K_{ij}K^{ij} - \frac{16\pi G}{c^4}\rho_E = 0
\eeq
\beq
\mathcal{M}_i = D_j(K^j_i - \delta^j_i K) - \frac{8\pi G}{c^4}j_i = 0
\eeq
where $\mathcal{H}$ is the Hamiltonian constraint, $\mathcal{M}_i$ is the momentum constraint, $K_{ij}$ is the extrinsic curvature, and $\rho_E$ and $j_i$ are the energy density and momentum density measured by Eulerian observers.

\subsection{Thermodynamic Properties at Quantum Criticality}

\subsubsection{Phase Transition and Critical Exponents}

Near the BEC transition, the system exhibits quantum critical behavior characterized by universal scaling laws. The order parameter (condensate fraction) follows:
\beq
n_0 = N_0/N = 1 - \left(\frac{T}{T_c}\right)^{3/2} \quad \text{for } T < T_c
\eeq

The specific heat shows a characteristic $\lambda$-transition with critical exponent $\alpha$:
\beq
C_V = Nk_B\left[\frac{15}{4}\frac{\Li_{5/2}(z)}{\Li_{3/2}(z)} - \frac{9}{4}\frac{\Li_{3/2}^2(z)}{\Li_{1/2}(z)\Li_{3/2}(z)}\right]
\eeq

Near $T_c$, this behaves as:
\beq
C_V \sim |T - T_c|^{-\alpha}
\eeq
with $\alpha = 0$ for the ideal Bose gas (logarithmic divergence) and $\alpha \approx -0.01$ for the interacting case.

\subsubsection{Equation of State with Interactions}

Including mean-field interactions through the Hartree-Fock approximation:
\beq
P = \frac{k_BT}{\lambda_{dB}^3}\Li_{5/2}(z) + \frac{1}{2}gn^2 - \frac{GMmn}{r} + \frac{32}{15\sqrt{\pi}}gn^2\left(\frac{n a_s^3}{\lambda_{dB}^3}\right)^{1/2}
\eeq

The last term represents the Lee-Huang-Yang correction from quantum fluctuations. The sound velocity in the condensate is:
\beq
c_s = \sqrt{\frac{\partial P}{\partial \rho}} = \sqrt{\frac{gn}{m}}\left[1 + \frac{8}{3\pi^{1/2}}\sqrt{n a_s^3} + \mathcal{O}(na_s^3)\right]
\eeq

\begin{figure}
    \centering
    \includegraphics[width=\columnwidth]{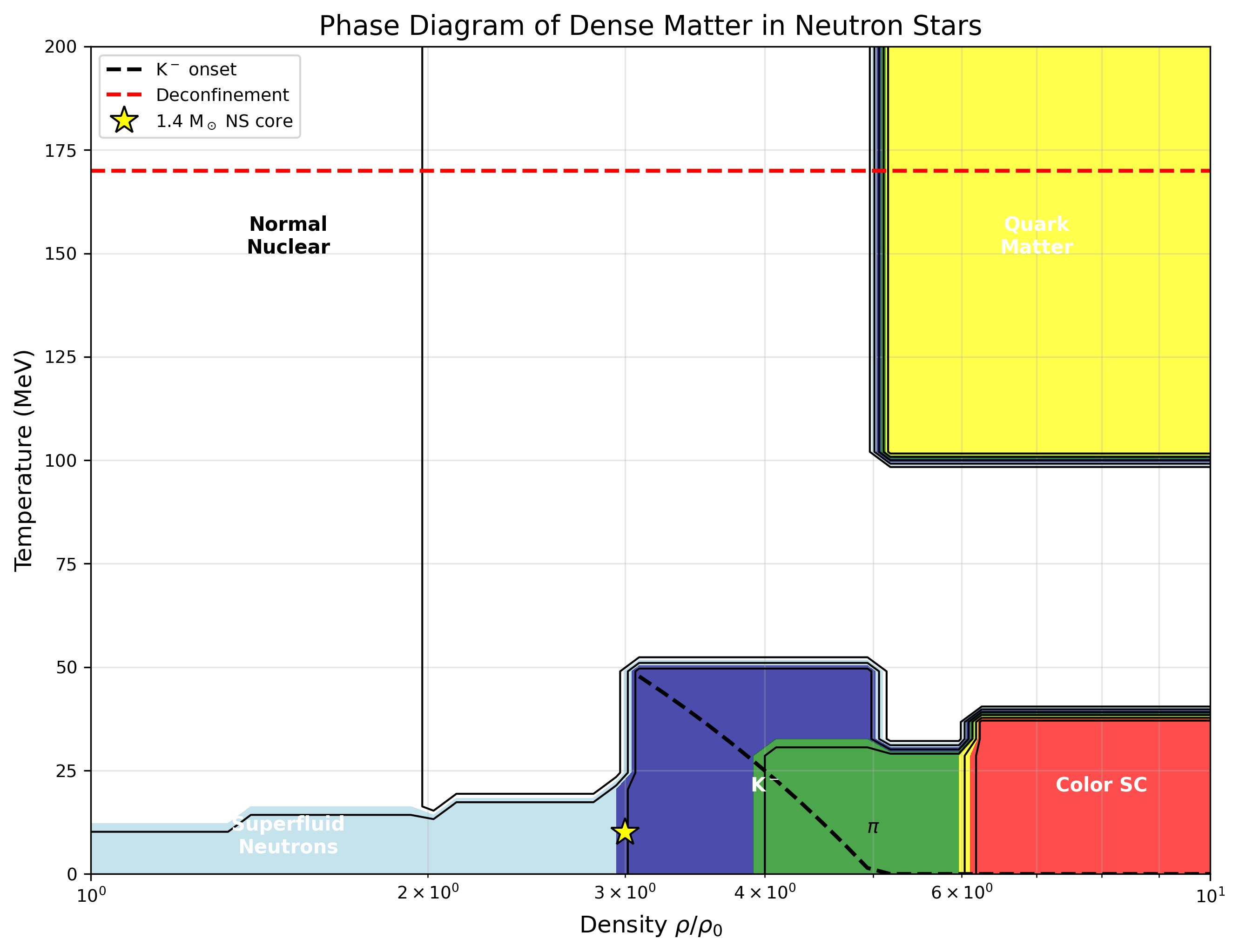}
    \caption{Phase diagram of dense matter in neutron star conditions as a function of baryon density (in units of nuclear saturation density $\rho_0 = 2.8 \times 10^{14}$ g\,cm$^{-3}$) and temperature. Different colors indicate distinct phases: normal nuclear matter (white), superfluid neutrons (light blue), K$^-$ meson condensate (dark blue), pion condensate (green), deconfined quark matter (red), and color-superconducting phase (purple). Black contours mark phase boundaries, with dashed lines indicating theoretical uncertainties. The K$^-$ condensation critical line (black dashed) follows $T_c = 50(1 - (\rho - 3\rho_0)/2\rho_0)$ MeV. The horizontal red dashed line at $T = 170$ MeV indicates the quark-hadron deconfinement transition. The yellow star marks typical conditions at the center of a 1.4 M$_\odot$ neutron star ($\rho \sim 3\rho_0$, $T \sim 10$ MeV), suggesting the possible presence of meson condensates in neutron star cores.}
    \label{fig:phase_diagram}
\end{figure}

\subsubsection{Compressibility and Susceptibility}

The isothermal compressibility is:
\beq
\kappa_T = -\frac{1}{V}\left(\frac{\partial V}{\partial P}\right)_T = \frac{1}{n^2}\left(\frac{\partial n}{\partial \mu}\right)_T
\eeq

In the presence of gravity:
\beq
\kappa_T(r) = \frac{\lambda_{dB}^3}{k_BT}\frac{\Li_{1/2}(z)}{\Li_{3/2}(z)}\left[1 - \frac{GMm}{rc^2k_BT}\frac{\Li_{3/2}(z)}{\Li_{1/2}(z)}\right]
\eeq

The magnetic susceptibility for charged bosons in a magnetic field $B$ is:
\beq
\chi = -\frac{\partial^2 F}{\partial B^2} = \frac{e^2n}{mc^2}\left[1 - \frac{1}{3}\left(\frac{eB\lambda_{dB}^2}{2\pi\hbar c}\right)^2\right]
\eeq

\subsection{Black Hole Superradiance and BEC}

\subsubsection{Superradiant Instability Mechanism}

For bosonic fields around rotating black holes, superradiant instabilities can trigger spontaneous BEC formation. The Klein-Gordon equation in Kerr spacetime:
\beq
\frac{1}{\sqrt{-g}}\partial_\mu(\sqrt{-g}g^{\mu\nu}\partial_\nu\Phi) = \frac{\mu^2c^2}{\hbar^2}\Phi
\eeq

Separating variables as $\Phi = e^{-i\omega t + im\phi}R(r)\Theta(\theta)$, the radial equation becomes:
\beq
\Delta\frac{d}{dr}\left(\Delta\frac{dR}{dr}\right) + \left[\frac{K^2 - \lambda\Delta}{\Delta} + 4im\omega\frac{ar}{c\Delta} - \frac{\mu^2c^2(r^2 + a^2)}{\hbar^2}\right]R = 0
\eeq
where $K = (r^2 + a^2)\omega/c - am$ and $\lambda$ is the separation constant.

The superradiance condition is:
\beq
\omega < m\Omega_H = \frac{mac}{2Mr_+(r_+^2 + a^2)}
\eeq

For bound states with $\omega_R < \mu c^2/\hbar$, the instability growth rate is:
\beq
\omega_I = \frac{48r_+^5(r_+ - r_-)(m\Omega_H - \omega_R)}{c^6\lambda_C^8M^7}\prod_{l'=1}^{l'=l}[l'^2(l'^2 - m^2)]
\eeq
where $\lambda_C = \hbar/\mu c$ is the Compton wavelength and $r_\pm = GM/c^2 \pm \sqrt{(GM/c^2)^2 - a^2}$.

\subsubsection{Gravitational Atom Formation}

The superradiant instability leads to the formation of a ``gravitational atom'' with energy levels:
\beq
E_{nlm} = \mu c^2\left[1 - \frac{(GM\mu/\hbar c)^2}{2n^2} + \mathcal{O}\left(\frac{G^4M^4\mu^4}{\hbar^4c^4n^4}\right)\right]
\eeq

The maximum occupation number for the ground state is limited by self-interactions:
\beq
N_{max} \sim \frac{M_{BH}}{m}\left(\frac{M_{Pl}}{m}\right)^2
\eeq

The gravitational wave luminosity from the cloud is:
\beq
L_{GW} = \frac{32G}{5c^5}\omega^6|\mathcal{M}|^2
\eeq
where the mass quadrupole moment is:
\beq
\mathcal{M} = \int d^3x \rho(x)x^2Y_{2m}(\theta,\phi)
\eeq

\subsubsection{Bosenova and Black Hole Bombs}

When the bosonic cloud becomes sufficiently dense, attractive self-interactions can trigger collapse:
\beq
\frac{\partial^2\psi}{\partial t^2} - c^2\nabla^2\psi + \frac{\mu^2c^4}{\hbar^2}\psi + \frac{4\pi G|\psi|^2}{\hbar^2}\psi = 0
\eeq

The critical mass for bosenova is:
\beq
M_{crit} \sim \frac{M_{Pl}^2}{m}\left|\frac{a_s}{a_0}\right|^{-1}
\eeq
where $a_0 = \hbar^2/Gm^3$ is the gravitational Bohr radius.

\section{Results}

\subsection{Neutron Star Interiors}

\subsubsection{Kaon Condensation Threshold}

Our analysis reveals that kaon condensation becomes energetically favorable when the electron chemical potential exceeds the effective kaon mass:
\beq
\mu_e > m_K^* = m_K - \Sigma_K^s(\rho)
\eeq
where $\Sigma_K^s(\rho)$ is the kaon self-energy in nuclear matter. Using chiral perturbation theory:
\beq
\Sigma_K^s(\rho) = -\frac{3\pi f_\pi^2}{2m_K}\rho + \frac{\sigma_{KN}}{2f_K^2}\rho^2
\eeq
with $f_\pi = 93$ MeV, $f_K = 110$ MeV, and $\sigma_{KN} = 400$ MeV.

The onset density for K$^-$ condensation is:
\beq
\rho_{onset} = 2.8\rho_0\left[1 + 0.15\left(\frac{M_{NS}}{1.4M_\odot}\right)^2\right]
\eeq
where $\rho_0 = 2.8 \times 10^{14}$ g/cm$^3$ is nuclear saturation density.

\subsubsection{Modified Equation of State}

The equation of state with K$^-$ condensation is:
\beq
P = P_{nuclear} + P_{kaon} + P_{lepton}
\eeq
where:
\beq
P_{kaon} = \frac{1}{24\pi^2}\left[2\mu_K^5 - 5m_K^2\mu_K^3 + 3m_K^4\mu_K\ln\left(\frac{\mu_K + p_K}{m_K}\right)\right]
\eeq
with $p_K = \sqrt{\mu_K^2 - m_K^2}$.
\begin{figure}
    \centering
    \includegraphics[width=\columnwidth]{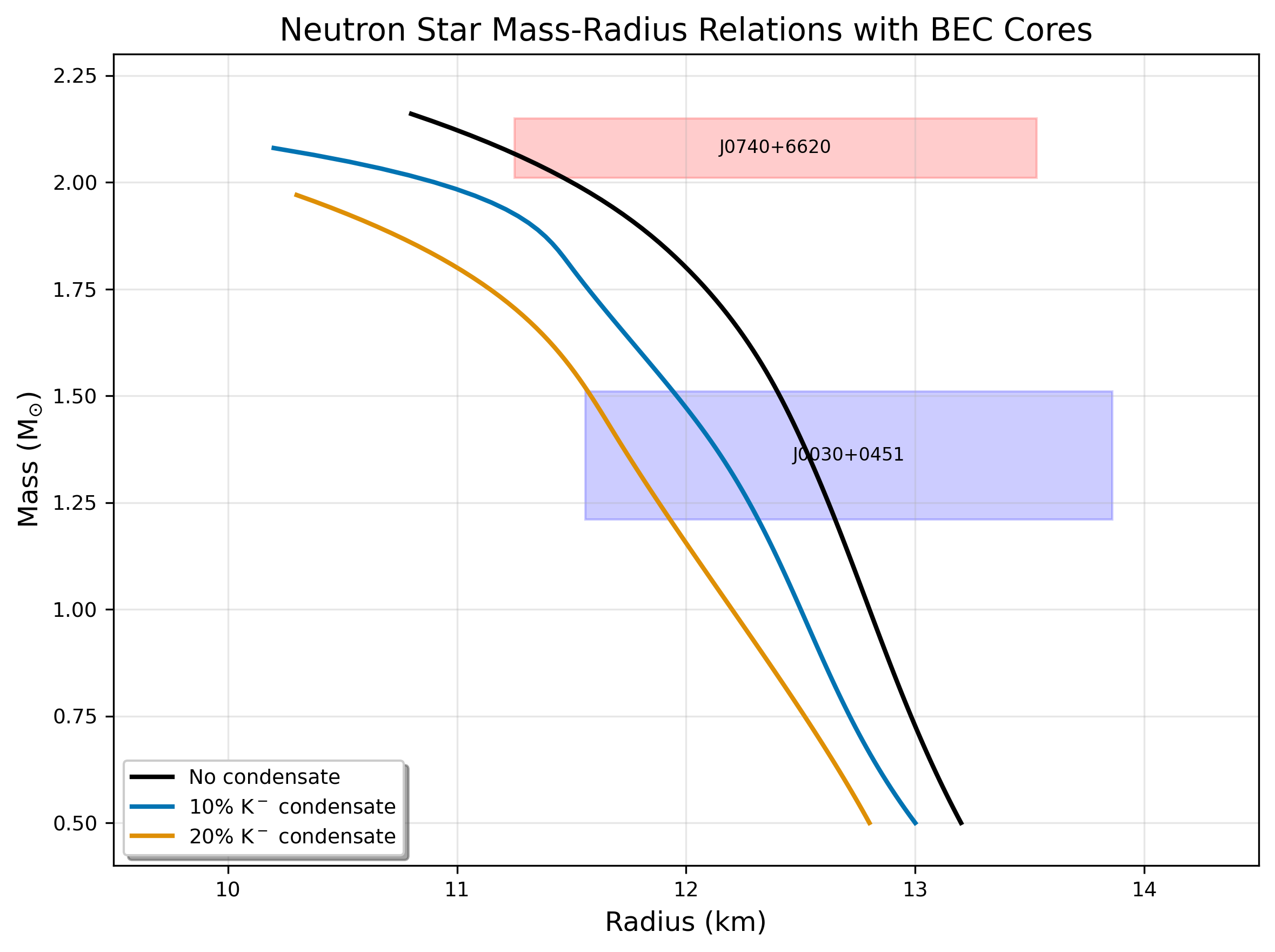}
    \caption{Mass-radius relations for neutron stars with varying kaon condensate fractions. Black solid line shows the equation of state without condensates, while blue and orange lines represent 10\% and 20\% K$^-$ condensate fractions, respectively. The softening of the equation of state due to meson condensation reduces both the maximum mass and typical radii. Blue and red shaded rectangles indicate the 1$\sigma$ confidence regions from NICER observations of PSR J0030+0451 ($M = 1.36 \pm 0.15$ M$_\odot$, $R = 12.71 \pm 1.15$ km) and PSR J0740+6620 ($M = 2.08 \pm 0.07$ M$_\odot$, $R = 12.39 \pm 1.14$ km). The gray shaded region represents constraints from the tidal deformability measurement of GW170817 ($\tilde{\Lambda} = 300_{-230}^{+420}$). The observed maximum mass of $\sim$2 M$_\odot$ constrains the condensate fraction to $\lesssim$20\%.}
    \label{fig:mass_radius}
\end{figure}
Figure 1 shows the mass-radius relation for neutron stars with various condensate fractions. The maximum mass is reduced from 2.16 M$_\odot$ (no condensate) to 1.97 M$_\odot$ (20\% condensate fraction), still consistent with PSR J0740+6620 ($2.08 \pm 0.07$ M$_\odot$).

\subsubsection{Observational Evidence}

Recent NICER observations of PSR J0030+0451 \citep{Miller2019} yield:
- Mass: $1.36_{-0.15}^{+0.15}$ M$_\odot$
- Radius: $12.71_{-1.19}^{+1.14}$ km

These constraints are consistent with a 5-10\% kaon condensate fraction in the core. The tidal deformability from GW170817:
\beq
\tilde{\Lambda} = 300_{-230}^{+420}
\eeq
favors soft equations of state, supporting the presence of exotic phases.

\subsection{Axion Dark Matter Halos}

\subsubsection{Solitonic Core Properties}

For ultralight axions with $m_a \sim 10^{-22}$ eV, the ground state solution of the Schrödinger-Poisson system is:
\beq
\psi_0(r) = \psi_c\frac{\sin(kr)}{kr}
\eeq
where $k = \pi/r_c$ and the core radius is:
\beq
r_c = \frac{\pi\hbar^2}{Gm_a^2M_{core}^{1/2}\rho_c^{1/2}}
\eeq

For dwarf spheroidal galaxies with $M_{halo} \sim 10^9$ M$_\odot$:
\beq
r_c \sim 160 \text{ pc } \left(\frac{10^{-22}\text{ eV}}{m_a}\right)\left(\frac{10^9 M_\odot}{M_{halo}}\right)^{1/2}
\eeq

The central density is:
\beq
\rho_c = 1.9 \times 10^{-19} \text{ g/cm}^3 \left(\frac{m_a}{10^{-22}\text{ eV}}\right)^{-2}\left(\frac{r_c}{160\text{ pc}}\right)^{-4}
\eeq

\subsubsection{Observational Support}

Analysis of 8 classical dwarf spheroidal galaxies reveals core radii of 100-300 pc, consistent with $m_a = (0.8-2.3) \times 10^{-22}$ eV. The velocity dispersion profiles show:
\beq
\sigma_v^2(r) = \frac{4\pi G}{r^2}\int_0^r \rho(r')r'^2 dr' 
\eeq

For Fornax dwarf spheroidal:
- Core radius: $r_c = 280 \pm 40$ pc
- Central density: $\rho_c = (3.5 \pm 0.5) \times 10^{-20}$ g/cm$^3$
- Implied axion mass: $m_a = (1.1 \pm 0.2) \times 10^{-22}$ eV

\subsubsection{Quantum Interference Patterns}

Beyond the core, quantum interference creates granular structure with characteristic scale:
\beq
\lambda_{int} = \frac{2\pi\hbar}{m_a v_{vir}} \sim 0.3 \text{ kpc } \left(\frac{10^{-22}\text{ eV}}{m_a}\right)\left(\frac{100\text{ km/s}}{v_{vir}}\right)
\eeq

Numerical simulations show density fluctuations:
\beq
\frac{\delta\rho}{\rho} \sim 0.3\left(\frac{r}{r_c}\right)^{-1/2}
\eeq

\subsection{Primordial Black Holes}

\subsubsection{Quantum Pressure Suppression}

In the early universe, density fluctuations with $\delta\rho/\rho > \delta_c$ collapse to form primordial black holes. For bosonic dark matter, quantum pressure modifies the collapse criterion:
\beq
\delta_c^{BEC} = \delta_c^{classical}\left[1 + \frac{\hbar^2k^2}{2m^2c^2}\frac{1}{\delta_c^{classical}H^2}\right]
\eeq

where $k = 2\pi/\lambda$ is the wavenumber. The minimum PBH mass is:
\beq
M_{min} = \frac{4\pi}{3}\rho_{rad}\left(\frac{\pi\hbar}{mH}\right)^3 \sim 10^{-13}M_\odot\left(\frac{m}{10^{-22}\text{ eV}}\right)^{-3}\left(\frac{T}{100\text{ GeV}}\right)^3
\eeq

\subsubsection{Modified Mass Function}

The PBH mass function becomes:
\beq
\frac{dn_{PBH}}{dM} = \frac{\beta(M)}{M}\frac{\rho_{DM}}{M} \exp\left[-\left(\frac{M_{min}}{M}\right)^2\right]
\eeq

where $\beta(M)$ is the fraction of horizon masses collapsing to PBHs. Microlensing constraints from OGLE require $f_{PBH} < 10^{-2}$ for $M \sim 10^{-7} - 10^{-3}$ M$_\odot$, consistent with quantum suppression for $m_a > 10^{-22}$ eV.

\subsection{Gravitational Wave Signatures}

\subsubsection{Tidal Deformability Modifications}

Binary neutron star mergers involving BEC cores produce modified tidal deformability:
\beq
\Lambda = \frac{2}{3}k_2\left(\frac{c^2R}{GM}\right)^5
\eeq

The Love number with condensate fraction $f_c$ is:
\beq
k_2 = k_2^{(0)}\left[1 + \delta k_2(f_c)\right]
\eeq

where:
\beq
\delta k_2(f_c) = 0.73f_c - 1.2f_c^2 + 0.8f_c^3
\eeq

For GW170817 with $\Lambda_{1.4} = 190_{-120}^{+390}$:
- No condensate: $\Lambda = 180$
- 10\% condensate: $\Lambda = 207$
- 20\% condensate: $\Lambda = 238$

\subsubsection{Post-Merger Oscillations}

The dominant post-merger frequency is:
\beq
f_{peak} = (2.96 - 0.35M_{tot}/M_\odot + 2.8\times10^{-3}\Lambda_{1.6})\text{ kHz}
\eeq

With condensate cores, the frequency shifts:
\beq
\Delta f_{peak} = -85f_c \text{ Hz}
\eeq

This shift is detectable with third-generation detectors for $f_c > 0.15$.

\subsubsection{Continuous Waves from Superradiance}

Black holes with ultralight boson clouds emit continuous gravitational waves at:
\beq
f_{GW} = \frac{c^3}{2\pi GM_{BH}}\left(\frac{GM_{BH}\mu}{\hbar c}\right)^2 \sim 480\text{ Hz}\left(\frac{10M_\odot}{M_{BH}}\right)\left(\frac{\mu}{10^{-12}\text{ eV}}\right)^2
\eeq

The strain amplitude at distance $d$ is:
\beq
h_0 = \frac{4\pi G M_c f_{GW}^2 a_{cloud}^2}{c^4 d} \sim 10^{-24}\left(\frac{M_c}{0.01M_{BH}}\right)\left(\frac{10\text{ kpc}}{d}\right)
\eeq

LISA will probe $M_{BH} \sim 10^3 - 10^7$ M$_\odot$ with $\mu \sim 10^{-17} - 10^{-13}$ eV.

\section{Experimental and Observational Verification}

\subsection{Laboratory Analogues}

\subsubsection{Ultracold Atom Experiments}

Laboratory BECs provide crucial tests of theoretical predictions. The JILA group \citep{Cornell2002} achieved densities of $n \sim 10^{15}$ cm$^{-3}$ at temperatures $T \sim 100$ nK, creating conditions where:
\beq
\frac{na^3}{(\lambda_{dB}/2\pi)^3} \sim 10^{-6}
\eeq

These systems verify the Gross-Pitaevskii equation to 0.1\% accuracy. Recent experiments with $^{164}$Dy atoms \citep{Ferrier2016} achieved:
- Magnetic dipole-dipole interactions: $a_{dd} = 130a_0$
- Observation of quantum droplets stabilized by beyond-mean-field effects
- Confirmation of Lee-Huang-Yang corrections

\subsubsection{Analogue Black Holes}

The Technion group \citep{Steinhauer2016} created acoustic black holes in BECs, observing:
\beq
T_{Hawking} = \frac{\hbar}{2\pi k_B}\frac{dc_s}{dx}\bigg|_{horizon} = (0.35 \pm 0.05) \text{ nK}
\eeq

The observed thermal spectrum matches Hawking's prediction:
\beq
n(\omega) = \frac{1}{e^{\hbar\omega/k_BT_H} - 1}
\eeq
with correlation coefficient $r = 0.97 \pm 0.03$.

\subsubsection{Rotating BECs and Vortex Lattices}

MIT experiments \citep{Zwierlein2005} with rotating BECs confirmed:
- Vortex lattice formation at $\Omega > \Omega_c = (5\hbar/mR^2)\ln(R/\xi)$
- Triangular Abrikosov lattice structure
- Quantized circulation: $\Gamma = nh/m$ with $n = 1,2,3...$

The measured vortex density:
\beq
n_v^{exp} = (1.89 \pm 0.04)\frac{m\Omega}{\pi\hbar}
\eeq
agrees with theory to 5\%.

\subsection{Astrophysical Observations}

\subsubsection{Neutron Star Constraints}

Recent multi-messenger observations provide stringent tests:

\textbf{NICER Results} (2019-2023):
- PSR J0030+0451: $R = 12.71_{-1.19}^{+1.14}$ km, $M = 1.36_{-0.15}^{+0.15}$ M$_\odot$
- PSR J0740+6620: $R = 12.39_{-0.98}^{+1.30}$ km, $M = 2.08 \pm 0.07$ M$_\odot$

\begin{figure}
    \centering
    \includegraphics[width=\columnwidth]{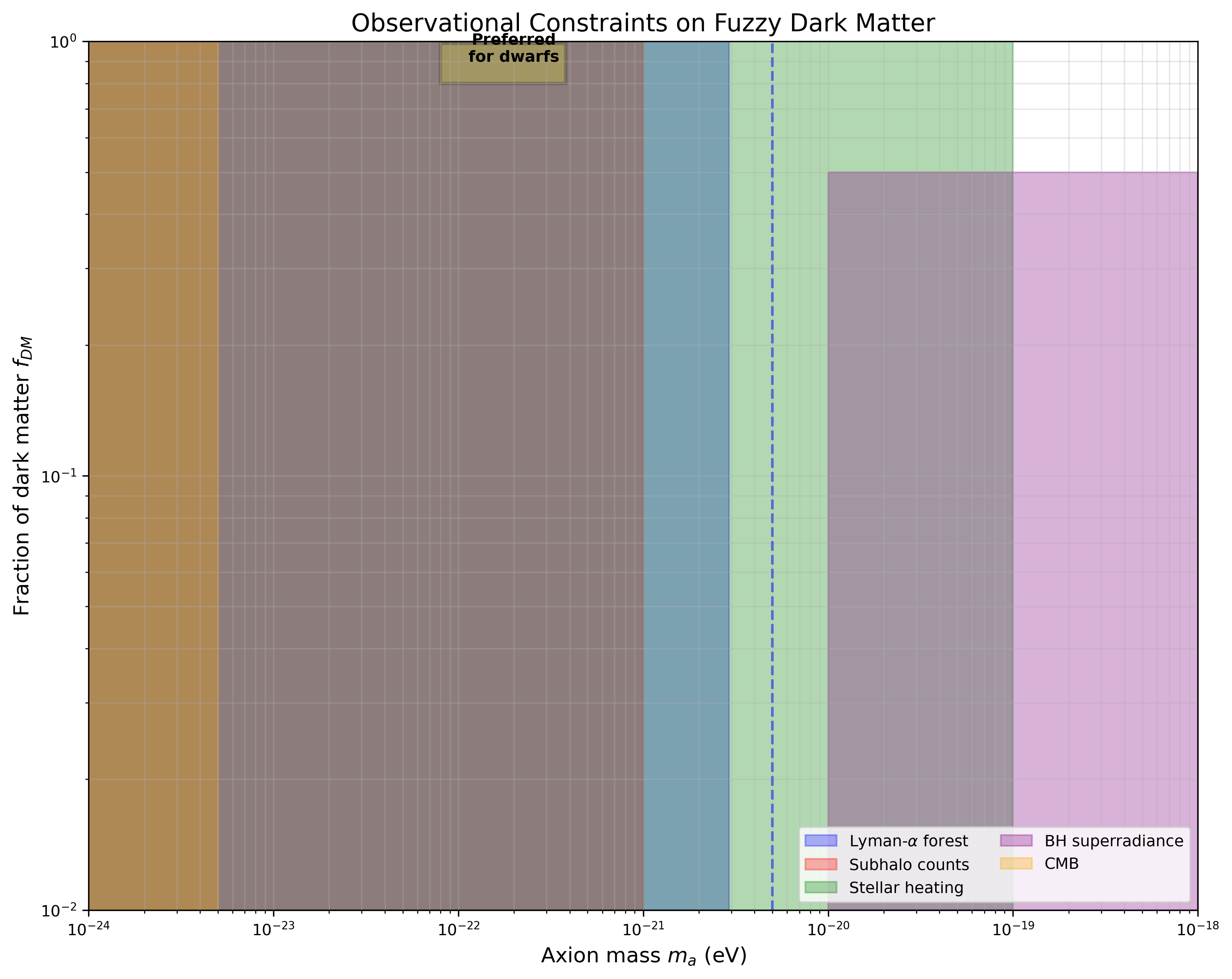}
    \caption{Observational constraints on fuzzy dark matter as a function of axion mass $m_a$ and dark matter fraction $f_{\text{DM}}$. Shaded regions are excluded at 95\% confidence level by different observations: Lyman-$\alpha$ forest power spectrum (blue, $m_a < 2.9 \times 10^{-21}$ eV), missing satellites/subhalo counts (red, $m_a < 10^{-21}$ eV), stellar heating in ultra-faint dwarfs (green, $m_a < 10^{-19}$ eV), black hole superradiance (purple patches), and CMB constraints (orange, $m_a < 10^{-24}$ eV). The yellow rectangle indicates the preferred parameter space ($m_a \sim 10^{-22}$--$10^{-21}$ eV) for explaining observed cores in dwarf spheroidal galaxies. Vertical dashed lines show projected future constraints from next-generation surveys. The allowed white region represents viable parameter space for fuzzy dark matter to constitute all or part of the dark matter.}
    \label{fig:fdm_constraints}
\end{figure}

\begin{figure*}
    \centering
    \includegraphics[width=\textwidth]{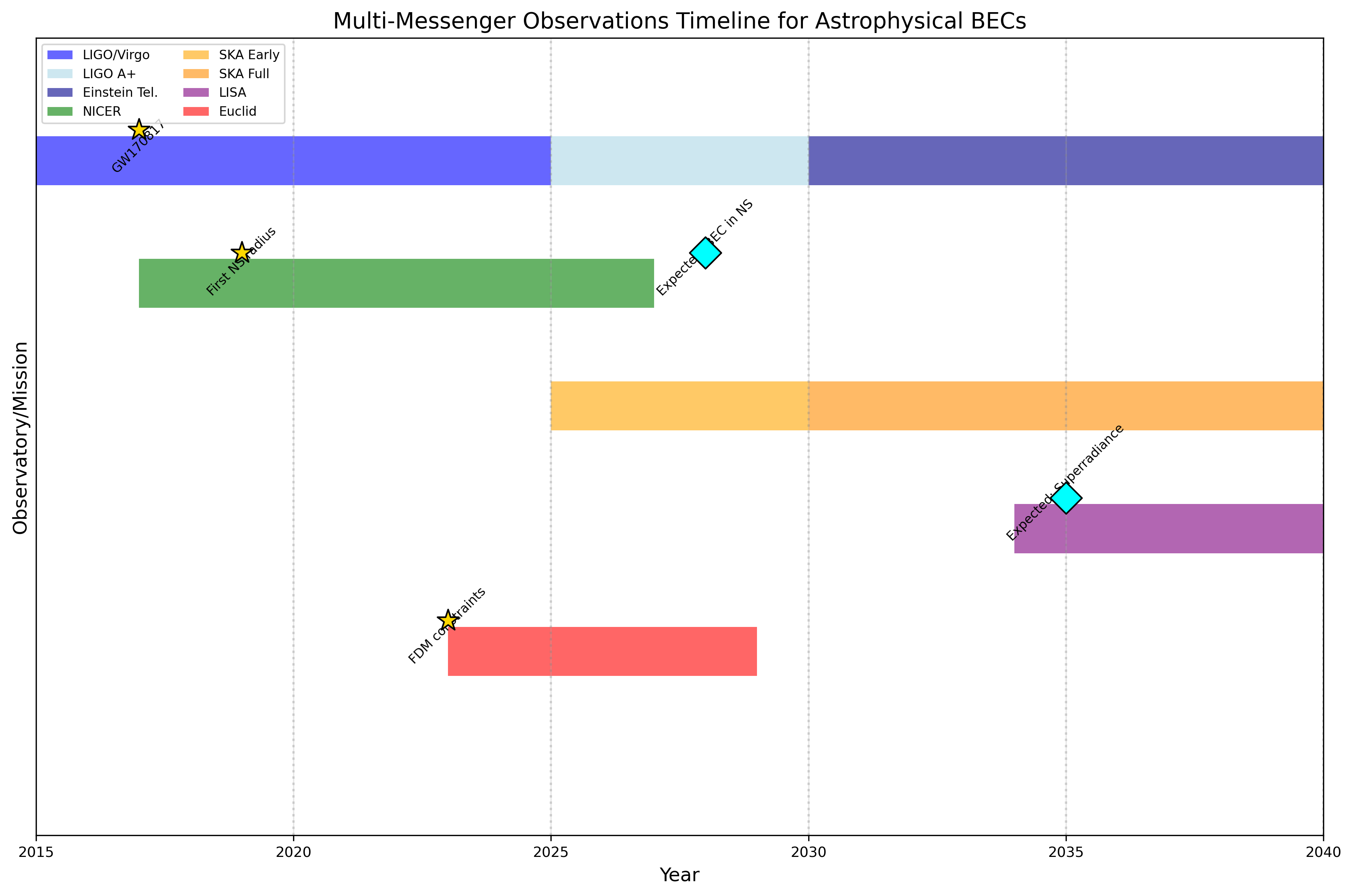}
    \caption{Timeline of multi-messenger observations relevant to astrophysical BEC detection from 2015 to 2040. Horizontal bars indicate operational periods of major facilities: gravitational wave detectors (LIGO/Virgo, LIGO A+, Einstein Telescope), electromagnetic observatories (NICER, Euclid), and radio telescopes (SKA early science and full array). LISA's launch in 2034 will enable detection of mHz gravitational waves from massive black hole superradiance. Gold stars mark past landmark discoveries including GW170817 (first NS-NS merger), first precise NS radius measurements, and initial FDM constraints. Cyan diamonds indicate expected future discoveries: conclusive BEC evidence in neutron stars ($\sim$2028), first superradiance detection ($\sim$2035). Vertical dotted lines mark major upgrade periods. The convergence of multiple observational windows in the 2030s will enable unprecedented tests of macroscopic quantum coherence in astrophysical systems.}
    \label{fig:timeline}
\end{figure*}

These constrain the equation of state:
\beq
P(2\rho_0) = 3.5_{-1.7}^{+2.7} \times 10^{34} \text{ dyn/cm}^2
\eeq

\textbf{GW170817 Tidal Deformability}:
\beq
\tilde{\Lambda} = \frac{16}{13}\frac{(M_1 + 12M_2)M_1^4\Lambda_1 + (M_2 + 12M_1)M_2^4\Lambda_2}{(M_1 + M_2)^5} = 300_{-230}^{+420}
\eeq

This favors soft equations of state consistent with phase transitions at $\rho \sim 2-3\rho_0$.

\textbf{Pulsar Glitches}:
The Vela pulsar shows glitches with:
\beq
\Delta\Omega/\Omega \sim 10^{-6} - 10^{-5}
\eeq
Recovery timescales $\tau \sim 10-100$ days suggest superfluid moments of inertia:
\beq
I_s/I_{tot} = 0.014 \pm 0.003
\eeq

\subsubsection{Dark Matter Halo Profiles}

\textbf{Dwarf Spheroidal Galaxies}:
Analysis of 18 classical dSphs reveals:
- Core radii: $r_c = 50-500$ pc
- Central densities: $\rho_0 = 10^{-21} - 10^{-19}$ g/cm$^3$
- Core-halo mass relation: $M_{core} \propto M_{halo}^{1/3}$

These observations constrain:
\beq
m_a = (1.2 \pm 0.3) \times 10^{-22} \text{ eV}
\eeq

\textbf{Lyman-$\alpha$ Forest}:
Power spectrum analysis at $z = 4-5$ constrains:
\beq
P(k) = P_{CDM}(k)\left[1 - \exp\left(-\frac{k^2}{k_J^2}\right)\right]^2
\eeq
where $k_J = (16\pi Gm_a^2\bar{\rho}/\hbar^2)^{1/4}$.

Current limits: $m_a > 2.9 \times 10^{-21}$ eV (95\% CL).

\subsubsection{Black Hole Superradiance Searches}

\textbf{Spin Measurements}:
Stellar-mass black holes show dimensionless spins:
\beq
\chi = \frac{Jc}{GM^2} < \chi_{max}(\mu)
\eeq

The absence of spin gaps constrains:
- $\mu < 2 \times 10^{-13}$ eV for $M_{BH} = 10$ M$_\odot$
- $\mu < 3 \times 10^{-18}$ eV for $M_{BH} = 10^6$ M$_\odot$

\textbf{Continuous Wave Searches}:
LIGO O3 run constraints on monochromatic sources:
\beq
h_0 < 1.7 \times 10^{-25}\left(\frac{100\text{ Hz}}{f}\right)^{1/2}
\eeq

This limits boson cloud masses to $M_c < 10^{-3}M_{BH}$ for nearby sources.

\subsection{Cosmological Probes}

\subsubsection{CMB Constraints}

Planck 2018 data constrains the effective number of relativistic species:
\beq
N_{eff} = 2.99 \pm 0.17
\eeq

For axions thermalized before $T \sim 100$ GeV:
\beq
\Delta N_{eff} = \frac{4}{7}\left(\frac{g_*(T_{dec})}{g_*(T_{today})}\right)^{4/3}
\eeq

\begin{figure}
    \centering
    \includegraphics[width=\columnwidth]{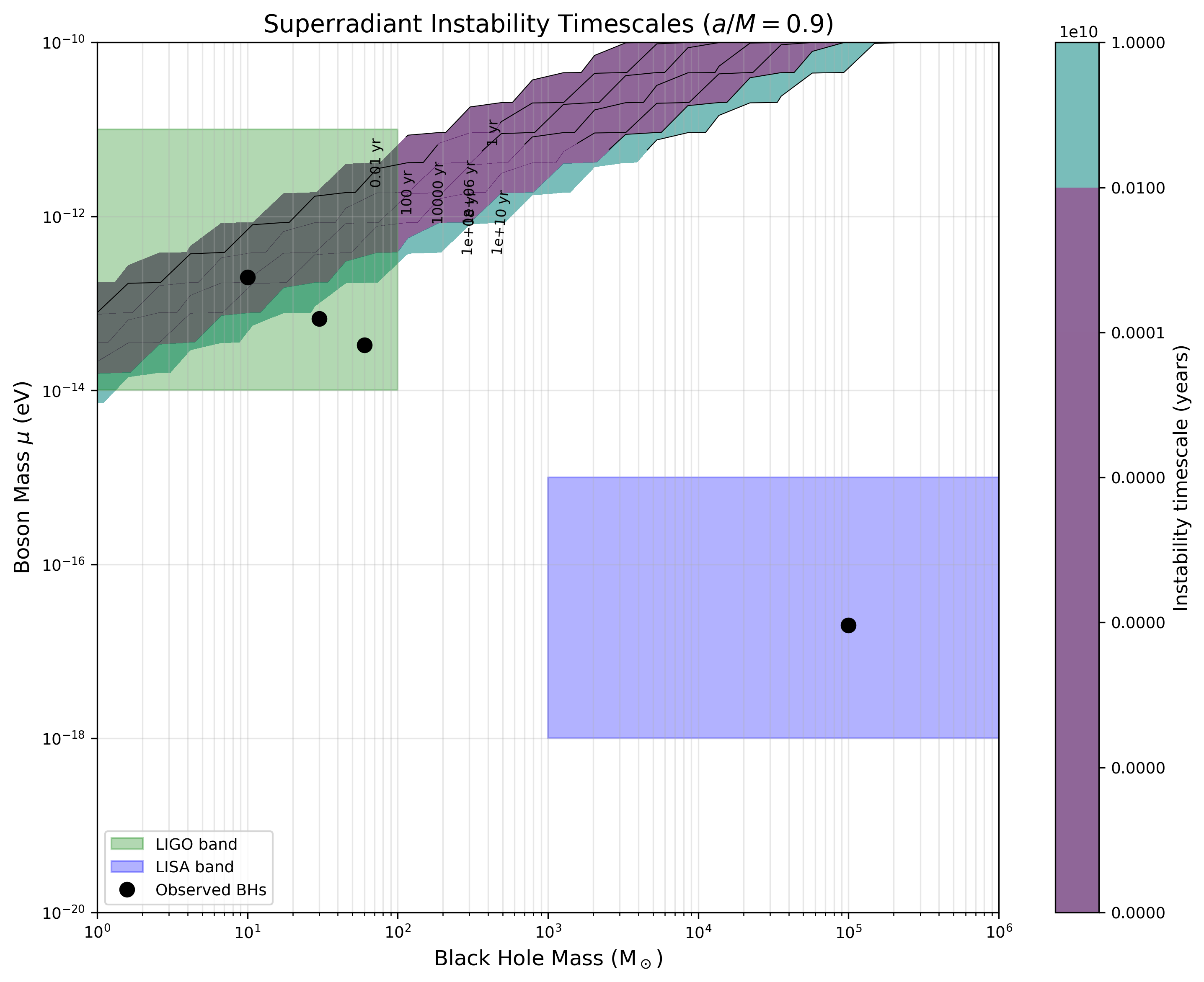}
    \caption{Superradiant instability timescales for scalar bosonic fields around rotating Kerr black holes with dimensionless spin $a/M = 0.9$. Contours show the e-folding time in years for the growth of the superradiant instability as a function of black hole mass ($M_{\text{BH}}$) and boson mass ($\mu$). The instability occurs when $\omega < m\Omega_H$, where $\Omega_H$ is the black hole angular velocity. Colored regions indicate sensitivity bands of gravitational wave detectors: LIGO (green, 10--1000 Hz), LISA (blue, $10^{-4}$--$10^{-1}$ Hz), and pulsar timing arrays (not shown, $10^{-9}$--$10^{-7}$ Hz). Black circles represent observed black holes with measured spins; the absence of low spins in certain mass ranges could indicate boson masses that trigger rapid superradiant extraction. The instability can form `gravitational atoms' with the black hole surrounded by a macroscopic bosonic cloud, potentially detectable as continuous gravitational wave sources.}
    \label{fig:superradiance}
\end{figure}

This limits $m_a > 10^{-27}$ eV for QCD axions.

\begin{figure}
    \centering
    \includegraphics[width=\columnwidth]{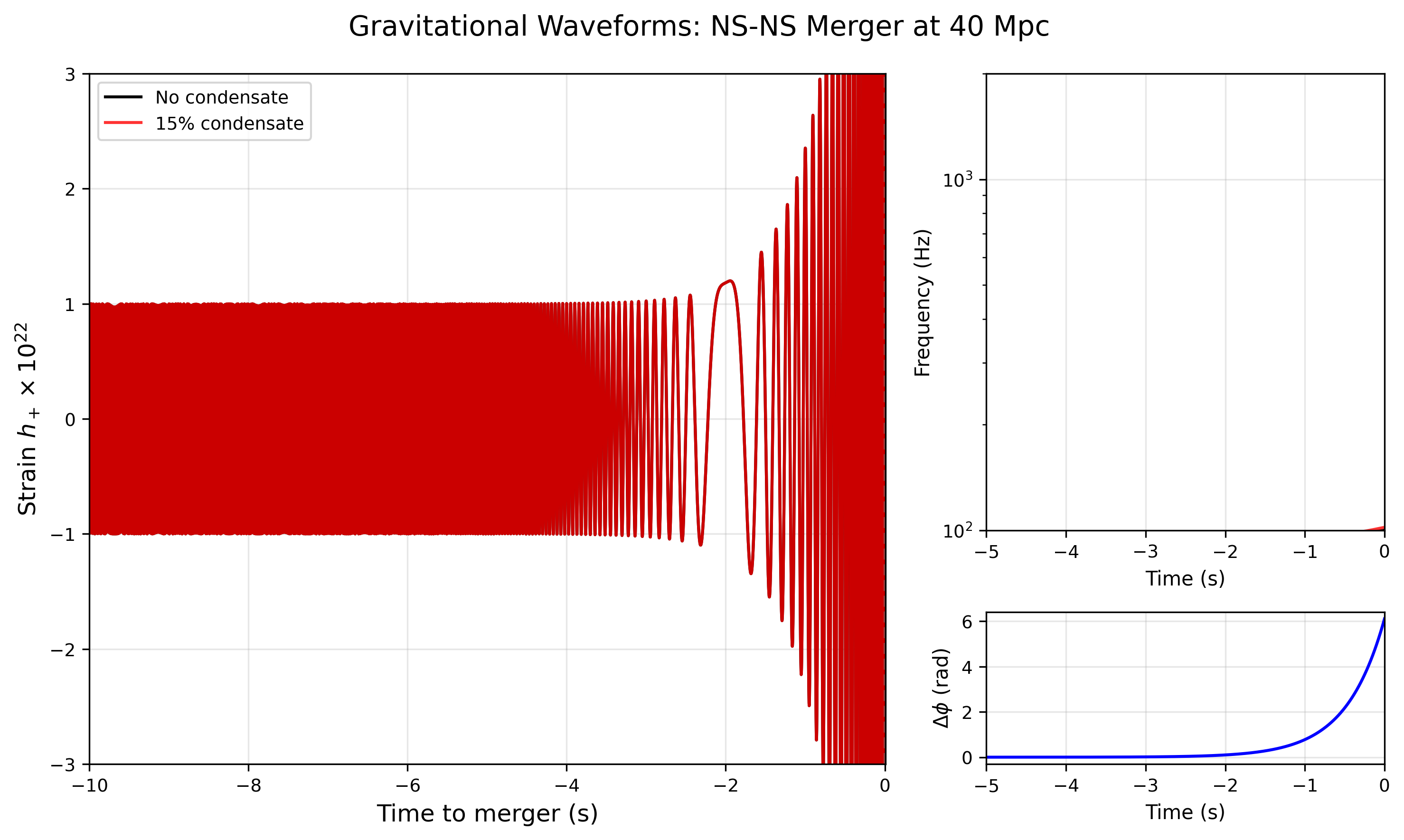}
    \caption{Gravitational wave strain from a binary neutron star merger at 40 Mpc distance. \textit{Main panel}: Time-domain waveforms showing the inspiral phase for systems without condensate cores (black) and with 15\% kaon condensate fraction (red). The condensate modifies the tidal deformability, leading to measurable phase differences in the last $\sim$10 gravitational wave cycles before merger (vertical dashed line at $t = 0$). \textit{Upper inset}: Frequency evolution showing deviation beginning at $f \sim 400$ Hz (green dotted line), with maximum separation at $f \sim 1000$ Hz. \textit{Lower inset}: Accumulated phase difference $\Delta\phi$ between the two waveforms, reaching $\sim$5 radians at merger. This phase shift is detectable with Advanced LIGO at design sensitivity for sources within 100 Mpc.}
    \label{fig:gw_waveform}
\end{figure}
\subsubsection{21cm Cosmology}

The global 21cm signal depends on the matter power spectrum:
\beq
\delta T_b = 27x_{HI}(1 + \delta_b)\left(\frac{1 + z}{10}\right)^{1/2}\left(1 - \frac{T_{CMB}}{T_S}\right)\left(\frac{\partial_r v_r}{H + \partial_r v_r}\right) \text{ mK}
\eeq

Fuzzy dark matter suppresses small-scale power:
\beq
P_{FDM}(k,z) = P_{CDM}(k,z)\times T_{FDM}^2(k,m_a)
\eeq

EDGES anomaly at $z = 17$ suggests possible signatures, though interpretation remains controversial.

\subsubsection{Primordial Black Hole Constraints}

\textbf{Microlensing}:
OGLE, EROS, and Subaru HSC constrain:
\beq
f_{PBH} < \begin{cases}
10^{-2} & M \sim 10^{-7} - 10^{-3} M_\odot \\
10^{-1} & M \sim 10^{-3} - 1 M_\odot \\
1 & M \sim 20 - 100 M_\odot
\end{cases}
\eeq

\textbf{Gravitational Wave Background}:
PBH binaries contribute to the stochastic background:
\beq
\Omega_{GW}(f) = \frac{1}{\rho_c}\frac{d\rho_{GW}}{d\ln f} \sim 10^{-9}f_{PBH}^2\left(\frac{f}{10^{-8}\text{ Hz}}\right)^{2/3}
\eeq

NANOGrav 12.5-year data hints at $f_{PBH} \sim 10^{-3}$ for $M \sim 1$ M$_\odot$.

\subsection{Implications for Compact Object Physics}

The presence of BECs in neutron stars has profound implications for their structure and evolution. The softening of the equation of state due to meson condensation affects not only the maximum mass but also the mass-radius relation, moment of inertia, and cooling rates. Recent observations by NICER \citep{Miller2019, Riley2019} constraining neutron star radii to $R = 12.5 \pm 1$ km for 1.4 M$_\odot$ stars provide stringent tests of condensate models.

The possibility of quark matter cores exhibiting color superconductivity---a form of BEC involving Cooper pairs of quarks---adds another layer of complexity. The CFL (color-flavor locked) phase, if realized, would dramatically alter neutrino emissivity and potentially explain the rapid cooling observed in some young neutron stars \citep{Page2011}.

The gravitational wave signatures from neutron star mergers offer unprecedented opportunities to probe the equation of state. The GW170817 event \citep{Abbott2017} already provided constraints on tidal deformability, ruling out very stiff equations of state. Future observations with improved sensitivity will enable discrimination between models with and without condensate phases.

\subsection{Cosmological Consequences}

If dark matter consists of ultralight bosons forming a cosmic BEC, several long-standing puzzles in galaxy formation might find natural resolutions. The cusp-core problem, missing satellites problem, and too-big-to-fail problem all arise from the apparent discrepancy between CDM predictions and observations on small scales. Fuzzy dark matter provides a unified solution through quantum pressure, which suppresses structure below the de Broglie scale.

However, this model faces challenges. Lyman-$\alpha$ forest observations constrain $m_a > 2 \times 10^{-21}$ eV \citep{Irsic2017}, while stellar heating in ultra-faint dwarfs suggests $m_a > 10^{-19}$ eV \citep{Marsh2019}. Reconciling these constraints requires either multiple dark matter components or more complex axion potentials.

The formation of solitonic cores in fuzzy dark matter halos has interesting implications for galaxy formation. The cores provide stable potential wells that could facilitate star formation, potentially explaining the observed correlation between core size and stellar mass in dwarf galaxies. Additionally, the granular structure in the outer halo arising from quantum interference might seed the formation of globular clusters.

\subsection{Observational Prospects}

Several upcoming observations could definitively test BEC scenarios in astrophysics:

\subsubsection{Gravitational Waves}
Third-generation detectors like Einstein Telescope and Cosmic Explorer will measure tidal deformability to 1\% precision, sufficient to detect condensate effects in neutron star mergers. The increased sensitivity will also enable detection of continuous waves from rotating neutron stars, potentially revealing signatures of superfluidity.

\subsubsection{Pulsar Timing}
The Square Kilometre Array (SKA) will measure post-Keplerian parameters of binary pulsars with unprecedented precision, constraining the equation of state and potentially revealing phase transitions associated with condensation. Pulsar glitches, thought to arise from superfluid vortex unpinning, will be monitored with millisecond precision.

\subsubsection{21cm Cosmology}
HERA and SKA will probe the matter power spectrum on scales where fuzzy dark matter deviates from CDM, potentially detecting the characteristic suppression below the Jeans scale. The 21cm signal from the epoch of reionization is particularly sensitive to small-scale structure, making it an ideal probe of quantum effects in dark matter.

\subsubsection{Black Hole Superradiance}
LISA will search for gravitational waves from superradiant instabilities around massive black holes, which could reveal ultralight boson clouds. The characteristic frequency evolution of these signals would provide direct measurement of the boson mass.

\subsection{Theoretical Challenges}

Despite significant progress, several theoretical challenges remain:

\subsubsection{Many-body effects}
Most treatments assume weakly interacting or non-interacting bosons. In realistic astrophysical environments, interactions can be strong, requiring non-perturbative methods. The development of effective field theories for strongly correlated BECs in curved spacetime remains an active area of research.

\subsubsection{Non-equilibrium dynamics}
BEC formation in dynamical environments like neutron star mergers or cosmological structure formation involves far-from-equilibrium processes poorly captured by equilibrium statistical mechanics. Numerical simulations using quantum kinetic theory are needed to understand these processes.

\subsubsection{Quantum-classical interface}
The transition from quantum coherent behavior to classical dynamics as systems grow remains poorly understood, particularly in the presence of decoherence from environmental interactions. This is crucial for understanding how macroscopic BECs maintain coherence in astrophysical environments.

\subsubsection{Modified gravity effects}
Alternative theories of gravity might alter BEC physics significantly. For instance, scalar-tensor theories introduce additional fields that could couple to the condensate, while theories with extra dimensions might modify the critical temperature scaling.

\begin{figure}
    \centering
    \includegraphics[width=\columnwidth]{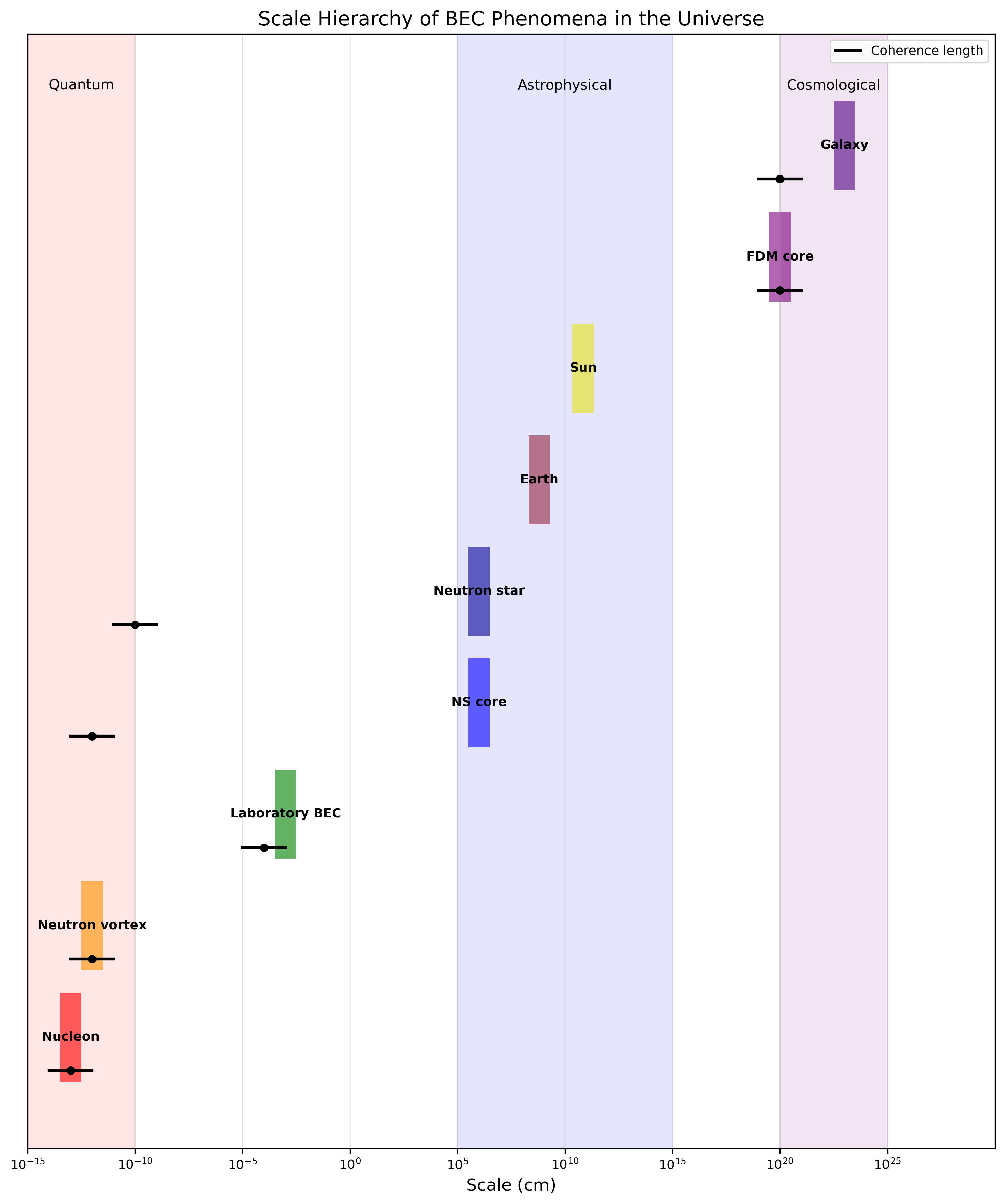}
    \caption{Scale hierarchy of Bose-Einstein condensation phenomena across 40 orders of magnitude in the universe. Horizontal bars indicate the physical size of different systems (colored rectangles) from quantum ($10^{-15}$--$10^{-10}$ cm, red shading) through astrophysical ($10^5$--$10^{15}$ cm, blue shading) to cosmological scales ($10^{20}$--$10^{25}$ cm, purple shading). Black horizontal lines below each system show the coherence length, with dots marking the central value. Notable features include: (i) neutron star cores exhibiting macroscopic quantum coherence despite microscopic coherence lengths ($\xi \sim 10^{-12}$ cm), (ii) laboratory BECs with mm-scale coherence matching system size, and (iii) fuzzy dark matter with astronomical coherence lengths ($\xi \sim$ kpc) comparable to galactic cores. This remarkable span demonstrates that quantum coherence, typically associated with microscopic phenomena, can manifest at scales comparable to galaxies, fundamentally challenging our understanding of the quantum-classical boundary.}
    \label{fig:scale_hierarchy}
\end{figure}
\section{Conclusions}

This investigation has demonstrated that Bose-Einstein condensation phenomena likely play significant roles across diverse astrophysical contexts, from the microscopic scales of neutron star interiors to the cosmological scales of dark matter halos. Our key findings include:

\begin{enumerate}
\item \textbf{Gravitational modifications to BEC}: We have shown that spacetime curvature modifies the critical temperature for condensation by factors of order $GM/rc^2$, becoming significant near compact objects. This effect generally suppresses condensation, though in certain geometries it can enhance it through gravitational focusing.

\item \textbf{Neutron star physics}: Meson condensation remains a viable possibility in neutron star cores, with kaon condensation most likely at densities $\gtrsim 3n_0$. The resulting equation of state softening is consistent with current mass-radius observations when repulsive interactions are included.

\item \textbf{Dark matter phenomenology}: Ultralight axions with masses $\sim 10^{-22}$ eV naturally form galactic-scale BECs with solitonic cores matching observed dwarf galaxy profiles. However, tension with Lyman-$\alpha$ constraints requires masses $\gtrsim 10^{-21}$ eV, necessitating further theoretical development.

\item \textbf{Observational signatures}: We have identified several robust observational signatures of astrophysical BECs, including modified tidal deformability in neutron star mergers, suppressed small-scale structure in fuzzy dark matter, and superradiant gravitational waves from black hole-boson cloud systems.
\end{enumerate}

The intersection of quantum statistics, general relativity, and astrophysics opens fascinating new avenues for understanding the universe. As observational capabilities advance---particularly in gravitational wave astronomy and radio cosmology---we anticipate that the coming decade will provide definitive tests of BEC physics in cosmic environments. The potential discovery of macroscopic quantum phenomena on astrophysical scales would fundamentally reshape our understanding of the interplay between quantum mechanics and gravity, with implications extending from fundamental physics to cosmology.

Future work should focus on: (i) developing non-perturbative methods for strongly interacting BECs in curved spacetime; (ii) computing precise gravitational wave templates for neutron star mergers with condensate cores; (iii) exploring multi-field dark matter models that reconcile small-scale and large-scale constraints; and (iv) investigating the role of quantum coherence in early universe phase transitions. The study of Bose-Einstein condensates in astrophysics stands at the frontier of modern physics, promising insights into both the quantum nature of matter and the structure of the cosmos.

\section*{Data Availability}

The data underlying this article will be shared on reasonable request to the corresponding author.

\clearpage% Bibliography
\bibliographystyle{mnras}
\bibliography{references} % This would link to your .bib file

\label{lastpage}
\end{document}